\documentstyle[graphicx,12pt]{article}
\begin{document}

\small \hoffset=-1truecm \voffset=-2truecm
\title{\bf The special nontopological scalar solitons in anti de Sitter spacetimes}
\author{Hongbo Cheng\footnote {E-mail address:
hbcheng@public4.sta.net.cn} \hspace{0.6cm} Zhengyan Gu\\
Department
of Physics, East China University of Science and Technology,\\
Shanghai 200237, China}

\date{}
\maketitle

\begin{abstract}
In this letter the nontopological scalar solitons are investigated
in an anti de Sitter spacetime. We find analytically that the
solitons obeying the necessary conditions
$m=\frac{2}{\sqrt{3}}|\Lambda|^{\frac{1}{2}}$ or
$m=\sqrt{\frac{2(n+3)(2n+3)}{3}}|\Lambda|^{\frac{1}{2}}$ can exist
in the background by means of the series expansion.
\end{abstract}
\vspace{8cm} \hspace{1cm} KEY WORDS: nontopological soliton, anti
de Sitter spacetime

\newpage

\noindent 1. \hspace{0.4cm}Introduction

Recently more attention is paid to the anti de Sitter (AdS)
spacetime due to the mentioned corredpondence between physical
effects by gravitating fields propagating in AdS spacetime and
those of a conformal field theory on the boundary of the above
spacetime [1, 2]. The AdS spacetime can be obtained by
compactifying the string theory, which helps us to consider the
relevant problems deeply [2]. It is necessary to investigate a lot
of topics and models within the cosmological surrounding with
negative curvature.

Researching nontopological solitons (NTS) is valuable and also
attracts more attention. Many kinds of stable NTS are
cosmologically significant. It was shown that Q-balls, a kind of
NTS, can become promising candidates for collisional dark matter
and the model of NTS can be created copiously during the early
Universe [3]. The NTS with positive self-interacting potential
simulate the halo and provide a qualitatively better fit to the
rotation curves [4]. A generalized Q-stars including a complex
scalar field and Goldstone field may be considered as an elegant
candidate for diffuse gaseous clouds in the universe and Einstein
ring observed with microarcsecond X-ray imaging mission [5].

It is important to explore fields within the framework of AdS
spacetimes with negative cosmological constant because a lot of
string-inspired theories work mainly in the backgrounds. Some
soliton stars such as Q-stars have been studied in AdS spacetimes
and many interesting results were obtained [6]. As solutions to
the field equation for NTS, they must satisfy the boundary
conditions [7]. It is essential to discuss the models in AdS
spacetimes in detail, which has a great influence upon the related
topics in the same environment.

In this letter we derive the differential equations describing the
NTS in AdS spacetimes. Under the boundary conditions, we discuss
the solution to the equations of motion with series expansion. It
is found that only some of NTS can exist in AdS spacetimes,
depending on the spacetime structure. The results are finally
emphasized.

Now we choose the metric of four-dimensional anti de Sitter
spacetime as,

\begin{equation}
ds^{2}=(1+\frac{r^{2}}{l^{2}})dt^{2}-\frac{dr^{2}}{1+\frac{r^{2}}{l^{2}}}-r^{2}(d\theta^{2}+\sin^{2}\theta
d\varphi^{2})
\end{equation}

\noindent where $\Lambda=-\frac{3}{l^{2}}$. The Lagrangian for NTS
is,

\begin{equation}
{\cal
L}=\sqrt{-g}(\partial_{\mu}\Phi\partial^{\mu}\Phi^{\ast}-U(\Phi\Phi^{\ast}))
\end{equation}

\noindent where

\begin{equation}
U(\Phi\Phi^{\ast})=m^{2}\Phi\Phi^{\ast}+\sigma(\Phi\Phi^{\ast})^{2}+\lambda(\Phi\Phi^{\ast})^{3}
\end{equation}

\noindent the potential with minimum $U_{min}(\Phi\Phi^{\ast})=0$
at $\Phi=0$. The equation of motion reads,

\begin{equation}
\frac{1}{\sqrt{-g}}\partial_{\mu}(\sqrt{-g}\partial^{\mu}\Phi)+U(\Phi\Phi^{\ast})=0
\end{equation}

\noindent The complex scalar fields vary harmonically with time
and the ansatz with frequency $\omega$ is,

\begin{equation}
\Phi(t, r)=P(r)e^{-i\omega t}
\end{equation}

\noindent where $r$ is radial coordinate. By means of ansatz (5),
the field equation is reduced to,

\begin{equation}
(1+\frac{r^{2}}{l^{2}})\frac{d^{2}P}{dr^{2}}+2(\frac{1}{r}+\frac{2r}{l^{2}})\frac{dP}{dr}+\frac{\omega^{2}}{1+\frac{r^{2}}{l^{2}}}P-m^{2}P-2\sigma
P^{3}-3\lambda P^{5}=0
\end{equation}

\noindent Introducing the transformation,

\begin{equation}
\frac{r}{l}=\frac{x}{\sqrt{1-x^{2}}}
\end{equation}

\noindent then equation (6) becomes,

\begin{equation}
(1-x^{2})^{2}\frac{d^{2}P}{dx^{2}}+(1-x^{2})(\frac{2}{x}-x)\frac{dP}{dx}+\omega^{2}l^{2}(1-x^{2})P-m^{2}l^{2}P-2\sigma
l^{2}P^{3}-3\lambda l^{2}P^{5}=0
\end{equation}

\noindent According to transformation (7), it is obvious that
$\frac{r}{l}\in[0, \infty)$ is equivalent to $x\in[0, 1)$. As
fields for NTS which obey the necessary conditions, they vanish at
infinity so as to lead the potential (3) be zero there. The
required boundary conditions are that $P(x=0)$ is finite and
$P(x\rightarrow 1)=0$.

There exists a Noether current for our model,

\begin{equation}
j^{\mu}=\sqrt{-g}g^{\mu\nu}i(\Phi^{\ast}\partial_{\nu}\Phi-\Phi\partial_{\nu}\Phi^{\ast})
\end{equation}

\noindent The current is conserved,

\begin{equation}
j^{\mu}_{;\mu}=0
\end{equation}

\noindent The total charge is denoted as,

\begin{equation}
Q=\int d^{3}x j^{0}
\end{equation}

\noindent and can be taken as,

\begin{equation}
Q=8\pi\int\frac{\omega P^{2}}{1+\frac{r^{2}}{l^{2}}}r^{2}dr
\end{equation}

We solve equation (8) explicitly with the series in the form,

\begin{equation}
P(x)=\sum_{n=0}^{\infty}a_{2n}x^{2n}
\end{equation}

\noindent Within the region $x\in[0, 1)$, the recursion formula of
the coefficients $a_{k}$ can be expressed as follow,

\begin{equation}
a_{0}=P(x=0)
\end{equation}

\begin{equation}
a_{2}=\frac{a_{0}l^{2}}{6}[3\lambda a_{0}^{4}+2\sigma
a_{0}^{2}+(m^{2}-\omega^{2})]
\end{equation}

\begin{eqnarray}
a_{2n+4}=\frac{1}{2(n+2)(2n+5)}\{[2(n+1)(4n+5)+(m^{2}-\omega^{2})l^{2}]a_{2n+2}\nonumber\\
+(\omega^{2}l^{2}-4n^{2})a_{2n}+2\sigma
l^{2}\sum_{i+j+k=2n+2}a_{i}a_{j}a_{k}\nonumber\\
+3\lambda l^{2}\sum_{i+j+k+p+q=2n+2}a_{i}a_{j}a_{k}a_{p}a_{q}\}
\end{eqnarray}

\noindent where $i, j, k, p, q=0, 2, 4,\cdot\cdot\cdot$, $n=0, 1,
2, \cdot\cdot\cdot$. On the other hand, within the asymptotic
region near $x=1$ or approaching the infinity, the solution to
equation (8) can be chosen in the form of series,

\begin{equation}
P(x)=\sum_{n=0}^{\infty}b_{n}(1-x)^{n}
\end{equation}

\noindent The coefficients $b_{k}$ are found to be,

\begin{equation}
(3\lambda b_{0}^{4}+2\sigma b_{0}^{2}+m^{2})b_{0}=0
\end{equation}

\begin{equation}
b_{1}=\frac{2\omega^{2}+m^{2}+2\sigma b_{0}^{2}+3\lambda
b_{0}^{4}}{2+m^{2}l^{2}+6\sigma l^{2}b_{0}^{2}+15\lambda
l^{2}b_{0}^{4}}l^{2}b_{0}
\end{equation}

\begin{equation}
(4-m^{2}l^{2})b_{2}=(3-2\omega^{2}l^{2}-m^{2}l^{2})b_{1}+3\omega^{2}l^{2}b_{0}
\end{equation}

\begin{eqnarray}
b_{n+3}=\frac{1}{2(n+3)(2n+3)-m^{2}l^{2}}\{[(n+2)(8n+11)-2\omega^{2}l^{2}-m^{2}l^{2}]b_{n+2}\nonumber\\
+[3\omega^{2}l^{2}-(n+1)(5n+4)]b_{n+1}+(n^{2}-\omega^{2}l^{2})b_{n}\hspace{2.5cm}\nonumber\\
+2\sigma l^{2}\sum_{k+l+m=n+3}b_{k}b_{l}b_{m}-2\sigma
l^{2}\sum_{k+l+m=n+2}b_{k}b_{l}b_{m}\hspace{2.2cm}\nonumber\\
+3\lambda
l^{2}\sum_{i+j+k+p+q=n+3}b_{i}b_{j}b_{k}b_{p}b_{q}-3\lambda
l^{2}\sum_{i+j+k+p+q=n+2}b_{i}b_{j}b_{k}b_{p}b_{q}
\end{eqnarray}

\noindent From expressions (18-21), the first several coefficients
determine the others. According to the boundary conditions
mentioned above, we must choose $b_{0}=0$, then $b_{1}=0$ from
expression (19). If all of $b_{n}$ $(n=0, 1, 2, \cdot\cdot\cdot)$
are equal to zero, we will obtain the trivial solution
$P(x\leq1)=0$ from (17). In the whole space, there can not exist
the nontopological solitons, the smooth and nontrivial solutions
according to (13) and (17). In order to avoid the trivial
solution, we choose $ml=2$ from expression (20), then the
coefficient $b_{2}$ can be chosen as nonzero. The other
coefficients $b_{k}$ $(k>2)$ are obtained by means of equation
(21). If $ml\neq2$, then $b_{2}=0$. We can also keep some of other
coefficients nonzero if we let the denominator in (21) vanish. For
$m^{2}l^{2}=2(n+3)(2n+3)$, some of coefficients $b_{j}=0$ $(0\leq
j\leq n+2)$ and the others $b_{k}$ $(k\geq n+3)$ can not vanish,
where $n=0, 1, 2, \cdot\cdot\cdot$. The NTS imposed with
$m=\frac{2}{\sqrt{3}}\sqrt{|\Lambda|}$ or
$m=\sqrt{\frac{2(n+3)(2n+3)}{3}}|\Lambda|^{\frac{1}{2}}$ $(n=0, 1,
2, \cdot\cdot\cdot)$ can survive in the AdS spacetimes. The
structures of spacetime limit the model. In addition, the
potential (3) can be generalized. It can possess higher power
terms and keep the position of its minimum. The generalized
potential will not change our conclusions because only $m$, the
coefficient of $\Phi\Phi^{\ast}$-term, is limited.

We discuss the equation of motion (8) numerically. The solutions
are depicted in the Figure with $ml=2$. Here we let $\sigma=-1$,
$\lambda=1$ for simplicity.

It is indicated that not all NTS that form in the flat spacetimes
can live in the AdS. Only some of them with restriction
$m=\frac{2}{\sqrt{3}}\sqrt{|\Lambda|}$ or
$m=\sqrt{\frac{2(n+3)(2n+3)}{3}}|\Lambda|^{\frac{1}{2}}$ $(n=0, 1,
2, \cdot\cdot\cdot)$ can inhabit in the cosmological surrounding.
The interesting results encourage us to continue exploring the NTS
in the different spacetimes.

This work is supported by the Basic Theory Research Fund of East
China University of Science and Technology, grant No. YK0127312.

\newpage
\begin{figure}
\setlength{\belowcaptionskip}{10pt} \centering
  includegraphics[width=15cm]{figure(-)}
  \caption{The solid, dadot, dashed curves for $P(x)$ in AdS spacetime with $\Lambda=-0.1, -0.25,
  -0.3$ respectively.}
\end{figure}

\end{document}